\documentstyle[sprocl,epsf,rotating]{article}

\newcommand{\lsim}{\raisebox{-0.13cm}{~\shortstack{$<$ \\[-0.07cm] $\sim$}}~}
\newcommand{\gsim}{\raisebox{-0.13cm}{~\shortstack{$>$ \\[-0.07cm] $\sim$}}~}
\newcommand{\tgb}{\mbox{tg}\beta}
\newcommand{\beq}{\begin{equation}}
\newcommand{\eeq}{\end{equation}}
\newcommand{\bea}{\begin{eqnarray}}
\newcommand{\eea}{\end{eqnarray}}
\newcommand{\non}{\nonumber}
\newcommand{\squ}{\tilde{Q}}
\newcommand{\MS}{$\overline{\rm MS}$}

\input{axodraw.sty}
\def\be{\begin{equation}}
\def\ee{\end{equation}}
\def\bea{\begin{eqnarray}}
\def\eea{\end{eqnarray}}

\begin{document}

\setcounter{page}{0}

\begin{large}
\begin{flushright}
CERN-TH/97-324 \\
hep-ph/9711407 \\
November 1997
\end{flushright}
\end{large}

\vspace*{1cm}

\renewcommand{\thefootnote}{\fnsymbol{footnote}}

\begin{center}
{\large \bf MSSM Higgs Boson Production at the LHC \footnote{Contribution
to the proceedings of the {\it International Workshop on Quantum Effects
in the MSSM}, 9--13 September 1997, Barcelona, Spain}}

\vspace*{1cm}

{\large Michael Spira}

\vspace*{1cm}

{\large \it CERN, Theory Division, CH-1211 Geneva 23, Switzerland}

\vspace*{2cm}


{\large \bf Abstract} 
\end{center}
 
\begin{large}
\noindent
The theoretical status of Higgs boson production at the LHC within the
minimal supersymmetric extension of the Standard Model is reviewed. The
focus will be on the evaluation of higher-order corrections to the
production cross sections and their phenomenological implications.

\vspace*{\fill}

\begin{flushleft}
CERN-TH/97-324 \\
hep-ph/9711407 \\
November 1997
\end{flushleft}
\end{large}

\thispagestyle{empty}

\newpage

\title{MSSM Higgs Boson Production at the LHC}

\author{MICHAEL SPIRA}

\address{CERN, Theory Division, CH-1211 Geneva 23, Switzerland \\
E-mail: Michael.Spira@cern.ch}


\maketitle\abstracts{
The theoretical status of Higgs boson production at the LHC within the
minimal supersymmetric extension of the Standard Model is reviewed. The
focus will be on the evaluation of higher-order corrections to the
production cross sections and their phenomenological implications.
}

\section{Introduction}
The search for Higgs bosons is one of the most important goals for
present and future experiments. Once a Higgs will be found, its properties have
to be investigated in order to distinguish between the Standard Model [SM] and
its extensions. Supersymmetry represents one of the most attractive extensions,
since it provides a natural solution to the hierarchy problem of the SM.
The minimal supersymmetric extension of the SM [MSSM] contains two isospin
doublets of Higgs fields, which are necessary to allow to introduce up and
down-quark
masses without breaking supersymmetry \cite{2higgs}. Moreover, they are
required to cancel
anomalies associated with Higgsinos, the supersymmetric partners of the
Higgs bosons. After electroweak symmetry breaking, one is left with a spectrum
of 5 elementary Higgs particles: two neutral CP-even
($h,H$), one neutral CP-odd ($A$) and two charged ($H^\pm$) Higgs bosons. The
MSSM Higgs sector can be described by two independent parameters at leading
order (LO), which are in general chosen as $\tgb = v_2/v_1$, i.e. the ratio of
the two vacuum expectation values of the scalar Higgs fields, and the
pseudoscalar mass $M_A$. Radiative corrections to the Higgs masses and
couplings are large \cite{mssmrad}, since the leading part $\epsilon$ grows
as the fourth power of the top-quark mass $m_t$:
\beq
\epsilon = \frac{3G_F}{\sqrt{2} \pi^2} \frac{m_t^4}{\sin^2\beta} \log\left(
1+\frac{\tilde m_s^2}{m_t^2} \right)
\eeq
where $G_F$ denotes the Fermi constant and $\tilde m_s$ the common squark mass.
They
shift the upper bound on the light scalar Higgs mass to $M_h\lsim 130$ GeV. The
Higgs couplings to top (bottom) quarks and gauge bosons are modified by SUSY
factors, which are collected in Table \ref{tb:susycoup}. The angle $\alpha$
denotes the mixing angle between the scalar Higgs particles $h,H$. An important
property of the SUSY couplings is the enhancement (suppression) of the bottom
(top) Yukawa coupling for increasing $\tgb$.
The direct search for the MSSM Higgs
particles at LEP yields lower limits $M_{h,H,A}\gsim 62.5$ GeV for the neutral
Higgs masses \cite{lep}.
\begin{table}[hbt]
\vspace*{-0.7cm}
\renewcommand{\arraystretch}{1.5}
\caption[]{\label{tb:susycoup}
\it Higgs couplings in the MSSM to fermions and gauge bosons [$V=W,Z$]
relative to SM couplings.}
\begin{center}
\begin{tabular}{|lc||ccc|} \hline
\multicolumn{2}{|c||}{$\phi$} & $g^\phi_t$ & $g^\phi_b$ &  $g^\phi_V$ \\
\hline \hline
SM~ & $H$ & 1 & 1 & 1 \\ \hline
MSSM~ & $h$ & $\cos\alpha/\sin\beta$ & $-\sin\alpha/\cos\beta$ &
$\sin(\beta-\alpha)$ \\
& $H$ & $\sin\alpha/\sin\beta$ & $\cos\alpha/\cos\beta$ &
$\cos(\beta-\alpha)$ \\
& $A$ & $ 1/\tgb$ & $\tgb$ & 0 \\ \hline
\end{tabular}
\renewcommand{\arraystretch}{1.2}
\end{center}
\end{table}

\begin{figure}[hbt]

\vspace*{0.3cm}
\hspace*{-0.2cm}
\begin{turn}{-90}%
\epsfxsize=4.0cm \epsfbox{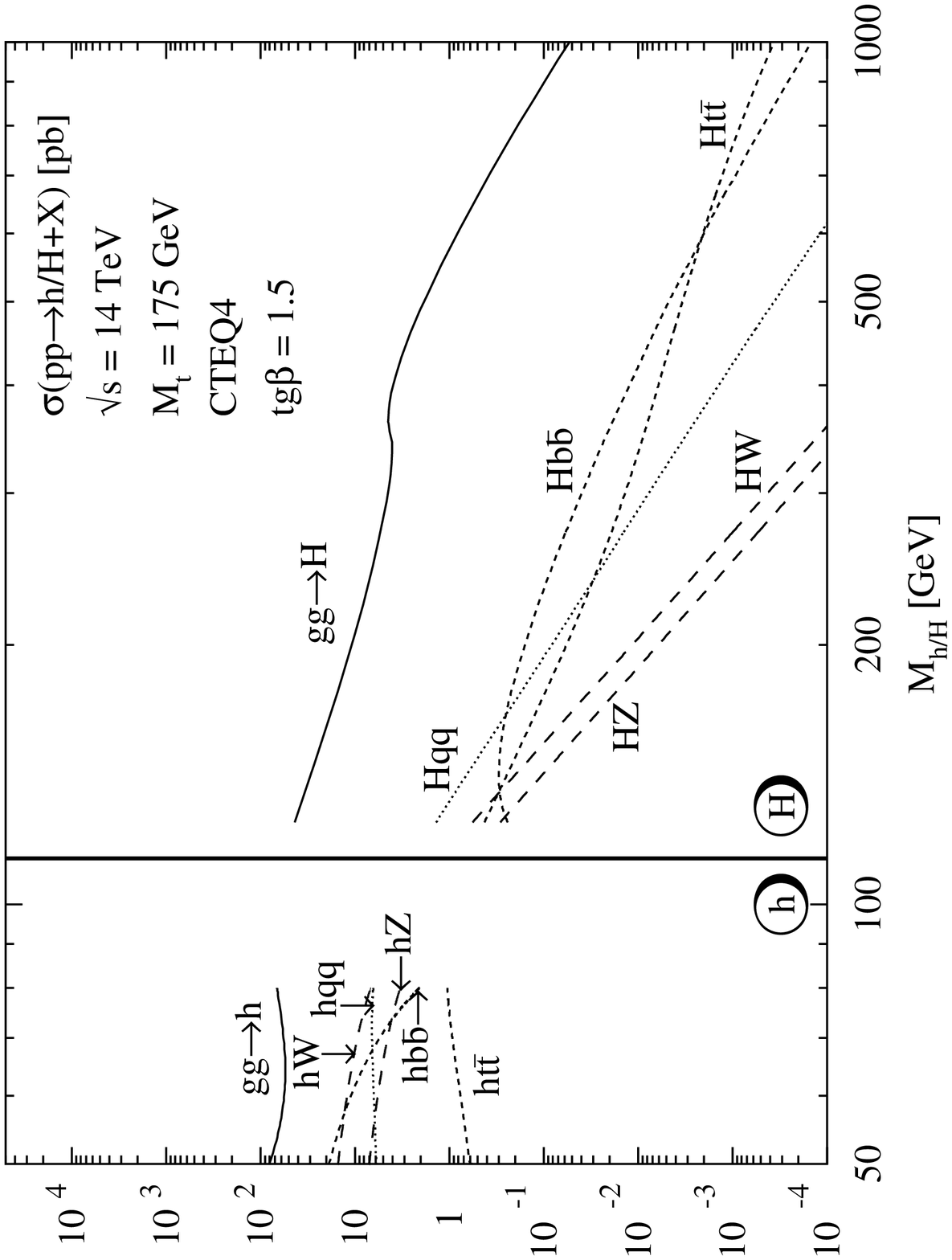}
\end{turn}
\vspace*{0.3cm}

\vspace*{-4.33cm}
\hspace*{5.7cm}
\begin{turn}{-90}%
\epsfxsize=4.0cm \epsfbox{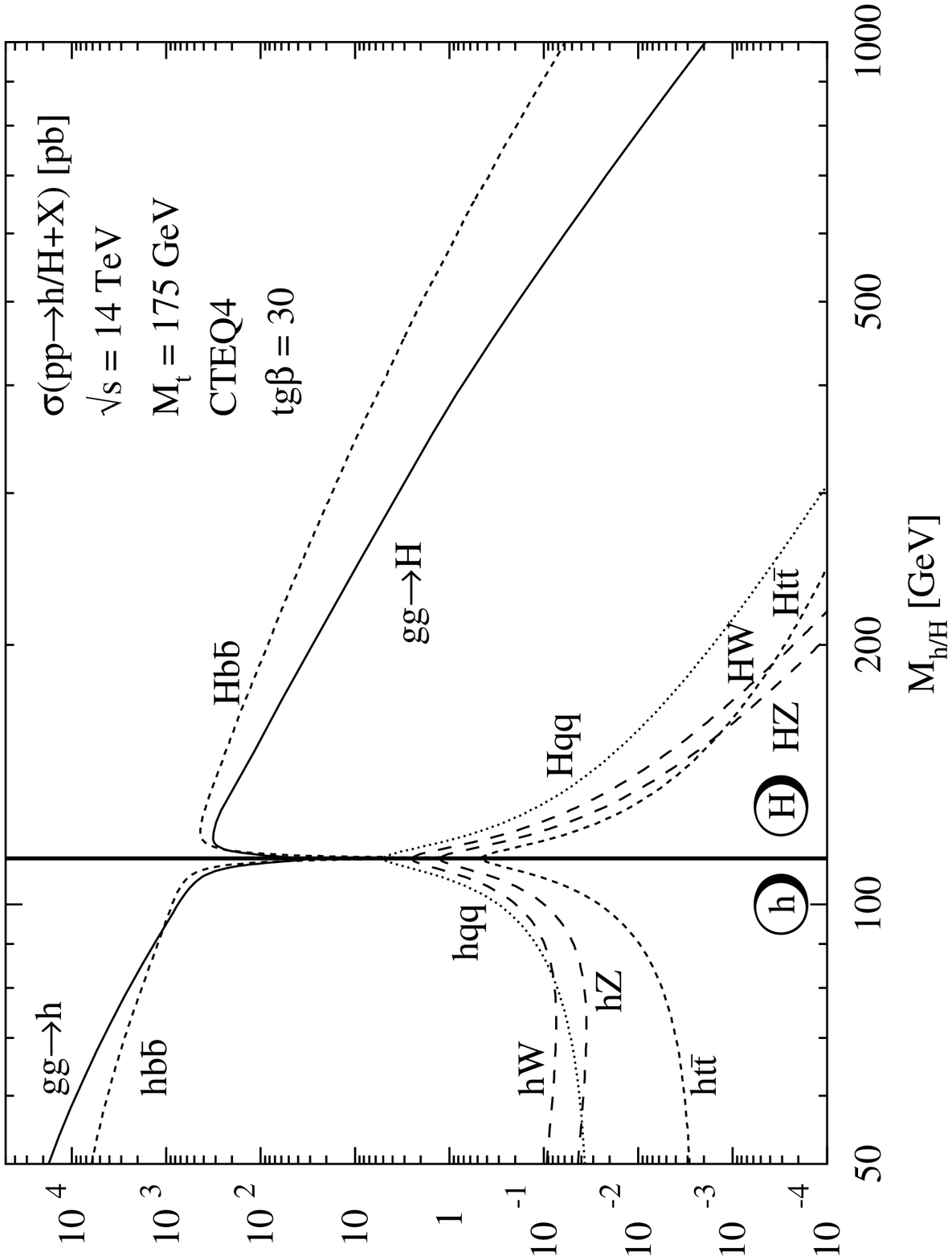}
\end{turn}
\vspace*{0.3cm}

\vspace*{0.0cm}
\hspace*{-0.2cm}
\begin{turn}{-90}%
\epsfxsize=4.0cm \epsfbox{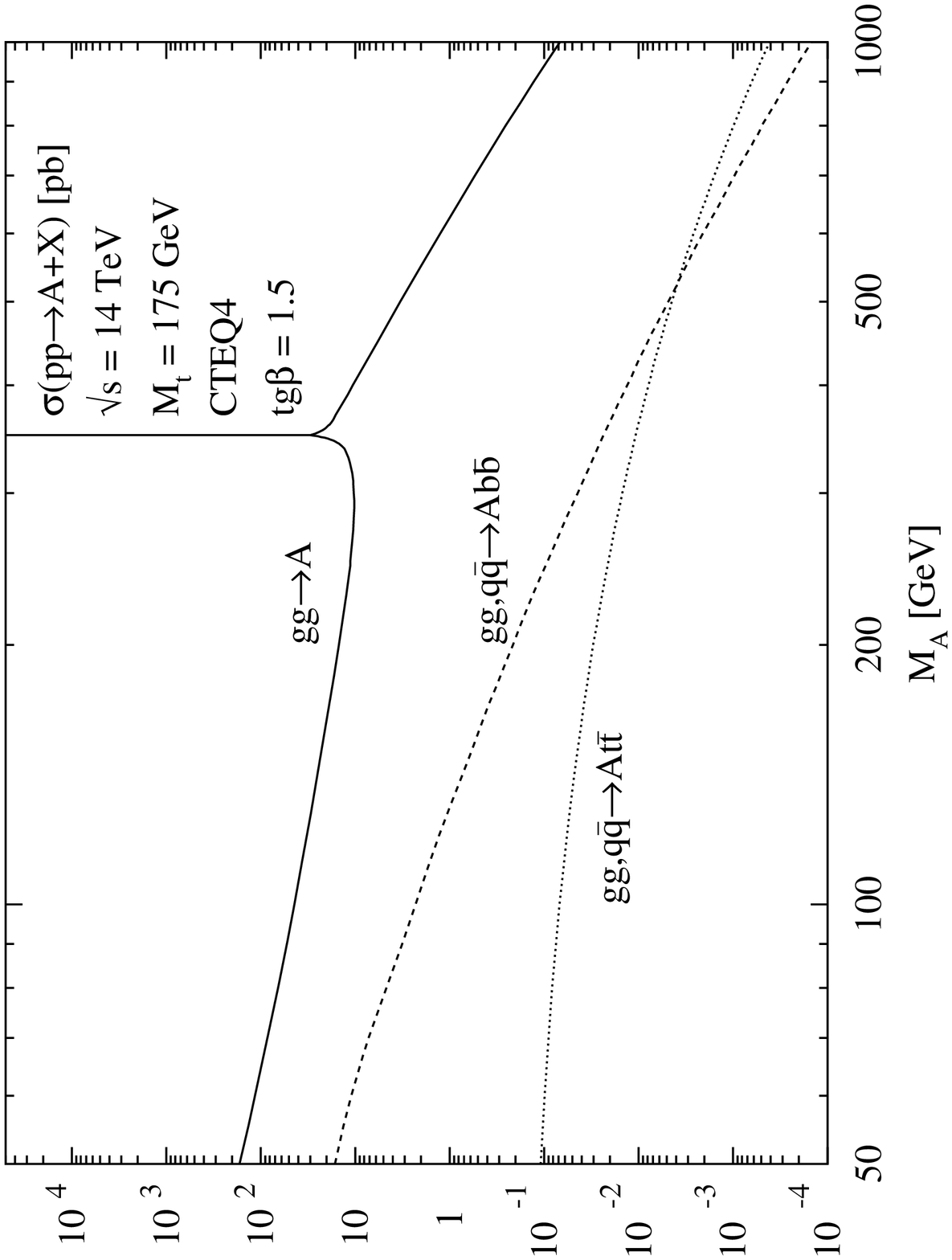}
\end{turn}
\vspace*{0.5cm}

\vspace*{-4.53cm}
\hspace*{5.7cm}
\begin{turn}{-90}%
\epsfxsize=4.0cm \epsfbox{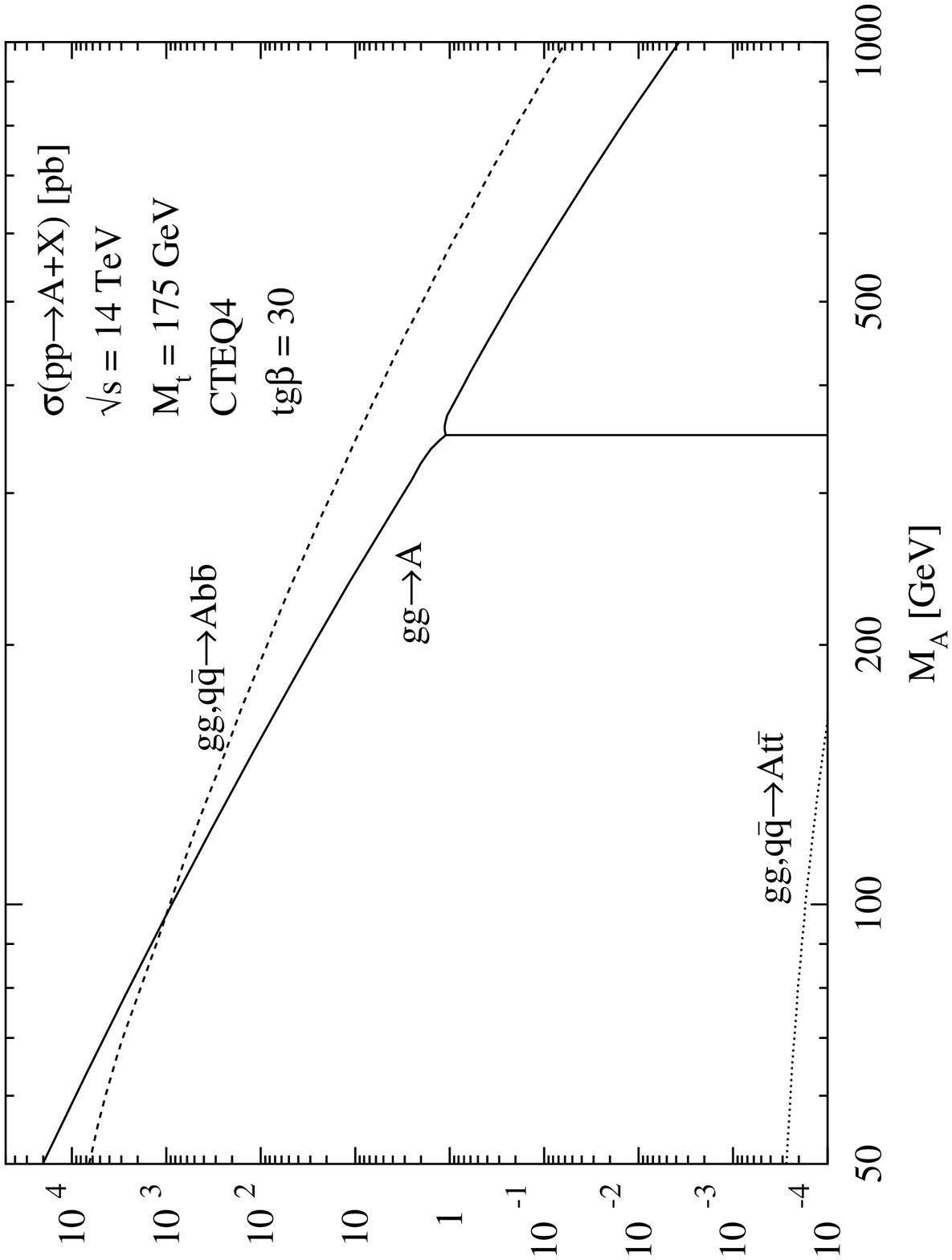}
\end{turn}
\vspace*{0.0cm}

\caption[]{\label{fg:prod} \it Neutral MSSM Higgs production cross
sections at the LHC [$\sqrt{s}=14$ TeV] for gluon fusion $gg\to \phi$,
vector-boson fusion $qq\to qqVV \to qqh/
qqH$, vector-boson bremsstrahlung $q\bar q\to V^* \to hV/HV$ and the associated
production $gg,q\bar q \to \phi b\bar b/ \phi t\bar t$ including all known
QCD corrections for $\tgb=1.5,30$.}
\vspace*{-0.5cm}
\end{figure}
At the LHC, the production of neutral Higgs bosons\footnote{Charged Higgs bosons
will be produced via radiation from a top quark \cite{tcb} or in pairs via the
Drell-Yan process \cite{dy} or gluon-gluon collisions \cite{ggcc}. They will
not be considered here.}
is dominated by gluon fusion $gg\to \phi$ [$\phi=h,H,A$] \cite{glufus}. Only
for large values of $\tgb$ does Higgs bremsstrahlung off
bottom quarks, $gg,q\bar q\to \phi b\bar b$, become dominant. This is shown
in Fig.~\ref{fg:prod}, in which the Higgs boson production
cross sections via the various mechanisms for the scalar and pseudoscalar
Higgs particles at the LHC are presented. For their discovery, several decay
modes must be
exploited in different regions of the MSSM parameter space \cite{atlas}:
$h\to\gamma\gamma$;
$H,A\to \tau^+\tau^-$; $H\to hh\to b\bar b \gamma\gamma$; $A\to ZH\to \ell^+
\ell^- b\bar b$; $H,A\to t\bar t$.

\section{$gg\to \phi$}
\subsection{Lowest order}
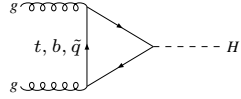
\begin{figure}[hbt]
\vspace*{-2.2cm}

\begin{center}
\SetScale{0.5}
\SetPFont{Times-Italic}{12}
\begin{picture}(180,100)(-50,15)

\Gluon(0,20)(50,20){-3}{5}
\Gluon(0,80)(50,80){3}{5}
\ArrowLine(50,20)(50,80)
\ArrowLine(50,80)(100,50)
\ArrowLine(100,50)(50,20)
\DashLine(100,50)(150,50){5}
\PText(160,46)(0)[l]{H}
\put(5,23){\scriptsize $t,b,\tilde q$}
\PText(-5,18)(0)[l]{g}
\PText(-5,78)(0)[l]{g}

\end{picture}  \\
\setlength{\unitlength}{1pt}
\caption[ ]{\label{fg:gghlodia} \it Diagrams contributing to $gg\to H$
at lowest order.}
\end{center}
\end{figure}
\begin{figure}[hbt]

\vspace*{-1.1cm}
\hspace*{2.5cm}
\epsfxsize=6cm \epsfbox{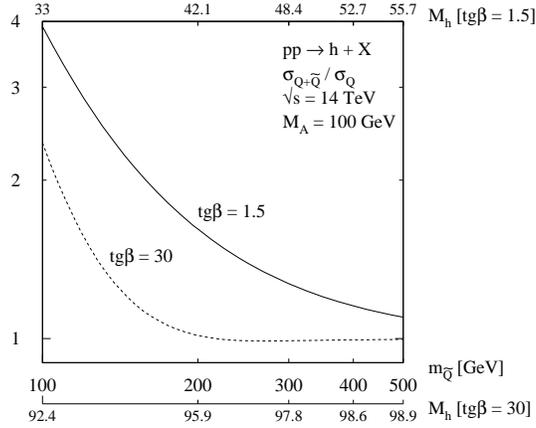}
\vspace*{-1.9cm}

\caption[]{\label{fg:squarkeffect} \it Ratio of the cross section
$\sigma(pp\to h + X)$ with and without squark loops as a function of the
common squark mass $m_{\squ}$ for two values of $\tgb=1.5, 30$, and
for $M_A=100$~GeV. The secondary axes present the corresponding light scalar
Higgs mass $M_h$.}
\vspace*{-0.5cm}
\end{figure}

At LO the gluon fusion mechanism is mediated by heavy top, bottom and squark
loops, see Fig.~\ref{fg:gghlodia}. The LO cross sections are given by
\cite{glufus,gghqcd}
\bea
\sigma_{LO}(pp\to \phi) & = & \sigma^\phi_0 \tau_\phi \frac{d{\cal L}^{gg}}
{d\tau_\phi} \label{eq:mssmgghlo} \\
\sigma^{h/H}_0 & = & \frac{G_{F}\alpha_{s}^{2}(\mu)}{288 \sqrt{2}\pi} \
\left| \sum_{Q} g_Q^{h/H} A_Q^{h/H} (\tau_{Q})
+ \sum_{\widetilde{Q}} g_{\widetilde{Q}}^{h/H} A_{\widetilde{Q}}^{h/H}
(\tau_{\widetilde{Q}}) \right|^{2} \nonumber \\
\sigma^A_0 & = & \frac{G_{F}\alpha_{s}^{2}(\mu)}{128 \sqrt{2}\pi} \
\left| \sum_{Q} g_Q^A A_Q^A (\tau_{Q}) \right|^{2} \nonumber
\end{eqnarray}
with the scaling variables $\tau_\phi = M^2_\phi/s$ and $\tau_{Q,\squ}=
4m_{Q,\squ}^2/M_\phi^2$ and the function
\begin{eqnarray}
f(\tau) & = & \left\{ \begin{array}{ll}
\displaystyle \arcsin^2 \frac{1}{\sqrt{\tau}} & \tau \ge 1 \\
\displaystyle - \frac{1}{4} \left[ \log \frac{1+\sqrt{1-\tau}}
{1-\sqrt{1-\tau}} - i\pi \right]^2 \hspace*{0.5cm} & \tau < 1
\end{array} \right.
\label{eq:ftau}
\end{eqnarray}
${\cal L}^{gg}$ denotes the gluon luminosity.
The top and bottom quarks as well as the third-generation squarks provide the
dominant contributions to the cross sections. Because of the behavior of the
SUSY couplings the top (bottom) contributions are suppressed (enhanced) for
large $\tgb$. The squark contributions are sizeable only if the squark mass
is below $\sim 400$ GeV, as can be seen in Fig.~\ref{fg:squarkeffect}, where
the ratio of the light scalar Higgs production cross sections with and without
the squark loops is shown as a function of the squark mass \cite{sqcd,habil}.

\subsection{QCD corrections}
\begin{figure}[hbt]
\vspace*{-1.9cm}

\begin{center}
\SetScale{0.5}
\SetPFont{Times-Italic}{12}
\begin{picture}(450,100)(-50,0)

\Gluon(0,20)(30,20){-3}{3}
\Gluon(0,80)(30,80){3}{3}
\Gluon(30,20)(60,20){-3}{3}
\Gluon(30,80)(60,80){3}{3}
\Gluon(30,20)(30,80){3}{5}
\ArrowLine(60,20)(60,80)
\ArrowLine(60,80)(90,50)
\ArrowLine(90,50)(60,20)
\DashLine(90,50)(120,50){5}
\PText(130,46)(0)[l]{H}
\PText(70,46)(0)[l]{t,b}
\PText(-5,18)(0)[l]{g}
\PText(-5,78)(0)[l]{g}
\PText(20,48)(0)[l]{g}

\Gluon(180,100)(210,100){3}{3}
\Gluon(210,100)(270,100){3}{6}
\Gluon(180,0)(210,0){-3}{3}
\Gluon(210,100)(210,60){3}{4}
\ArrowLine(210,0)(210,60)
\ArrowLine(210,60)(240,30)
\ArrowLine(240,30)(210,0)
\DashLine(240,30)(270,30){5}
\PText(280,26)(0)[l]{H}
\PText(220,26)(0)[l]{t,b}
\PText(170,-2)(0)[l]{g}
\PText(170,98)(0)[l]{g}
\PText(200,78)(0)[l]{g}
\PText(280,98)(0)[l]{g}

\ArrowLine(330,100)(360,100)
\ArrowLine(360,100)(420,100)
\Gluon(330,0)(360,0){-3}{3}
\Gluon(360,100)(360,60){3}{4}
\ArrowLine(360,0)(360,60)
\ArrowLine(360,60)(390,30)
\ArrowLine(390,30)(360,0)
\DashLine(390,30)(420,30){5}
\PText(430,26)(0)[l]{H}
\PText(370,26)(0)[l]{t,b}
\PText(320,-2)(0)[l]{g}
\PText(320,98)(0)[l]{q}
\PText(430,98)(0)[l]{q}
\PText(350,78)(0)[l]{g}

\end{picture}  \\
\setlength{\unitlength}{1pt}
\caption[ ]{\label{fg:gghqcddia} \it Typical diagrams contributing to the
virtual and real QCD corrections to $gg\to H$.}
\end{center}
\vspace*{-1.0cm}
\end{figure}
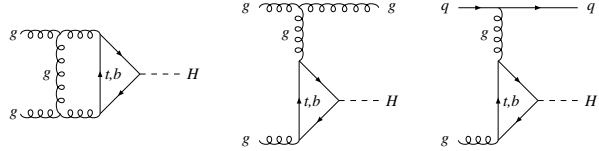
\begin{figure}[t]

\vspace*{-0.3cm}
\hspace*{-3.5cm}
\begin{turn}{-90}%
\epsfxsize=9cm \epsfbox{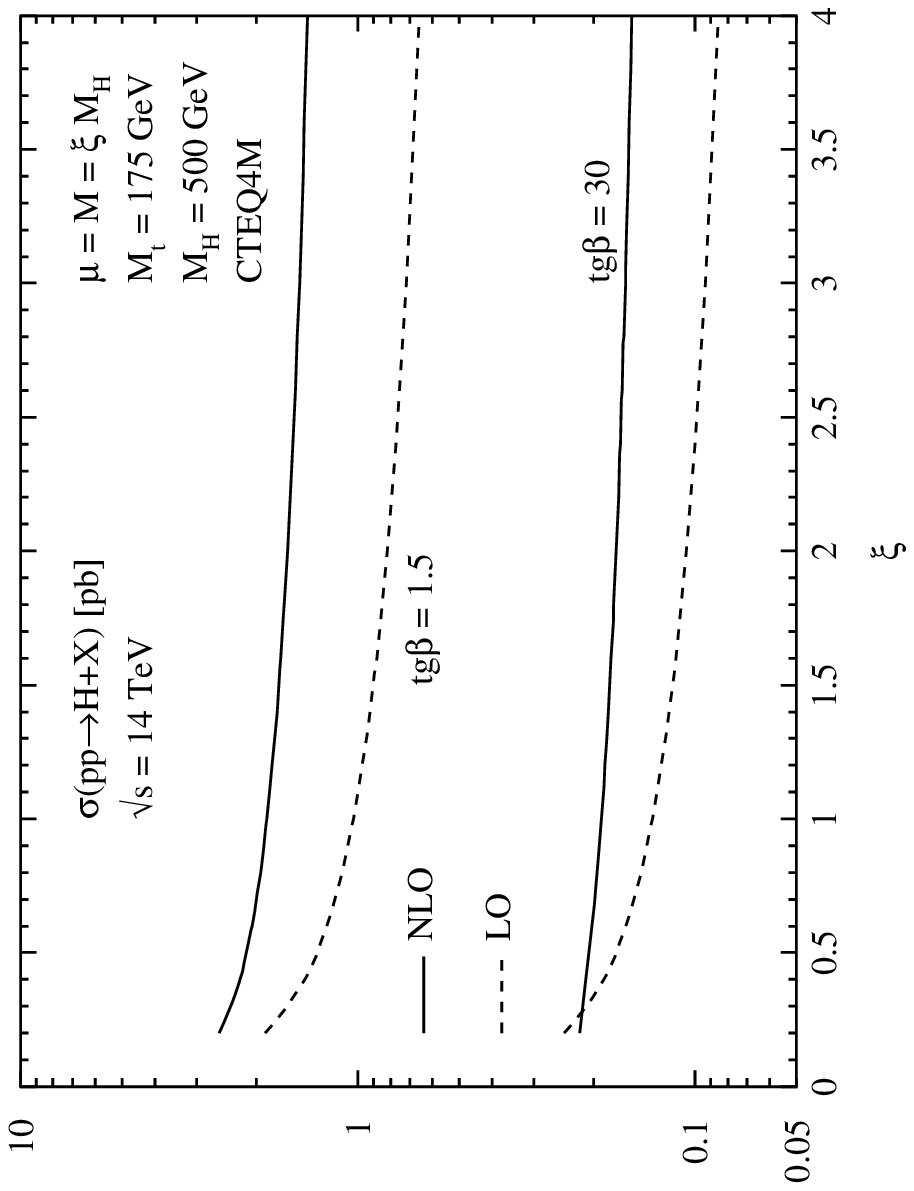}
\end{turn}
\vspace*{0.5cm}

\vspace*{-13.58cm}
\hspace*{2.5cm}
\begin{turn}{-90}%
\epsfxsize=9cm \epsfbox{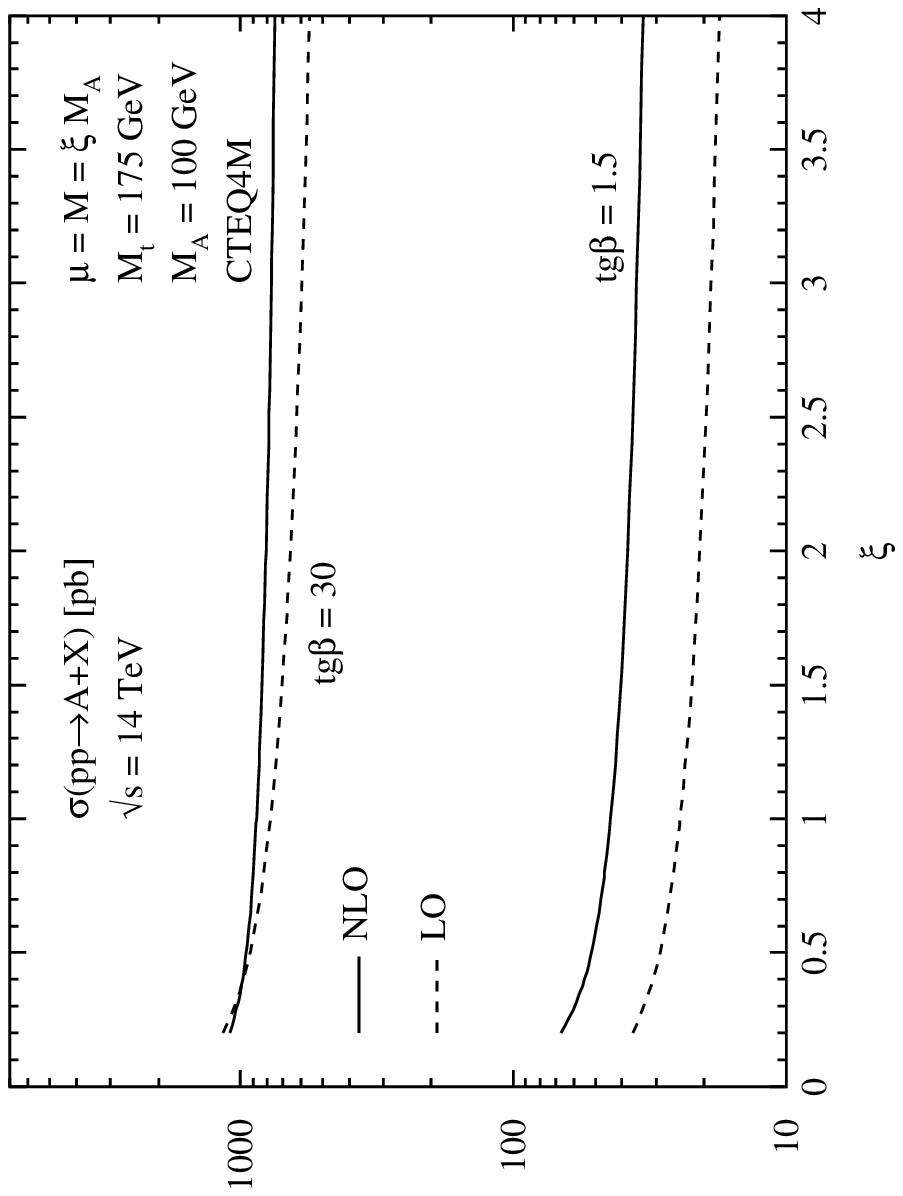}
\end{turn}
\vspace*{-0.8cm}

\caption[]{\label{fg:mssmgghscale} \it The renormalization and factorization
scale dependence of the Higgs production cross section at lowest and
next-to-leading order for two different Higgs bosons $H,A$ with masses $M_H =
500$ GeV and $M_A=100$ GeV and two values of $\tgb=1.5,30$.}
\vspace*{-0.5cm}
\end{figure}

\paragraph{\it Quark loops.}
\begin{figure}[t]

\vspace*{-0.3cm}
\hspace*{-3.5cm}
\begin{turn}{-90}%
\epsfxsize=9cm \epsfbox{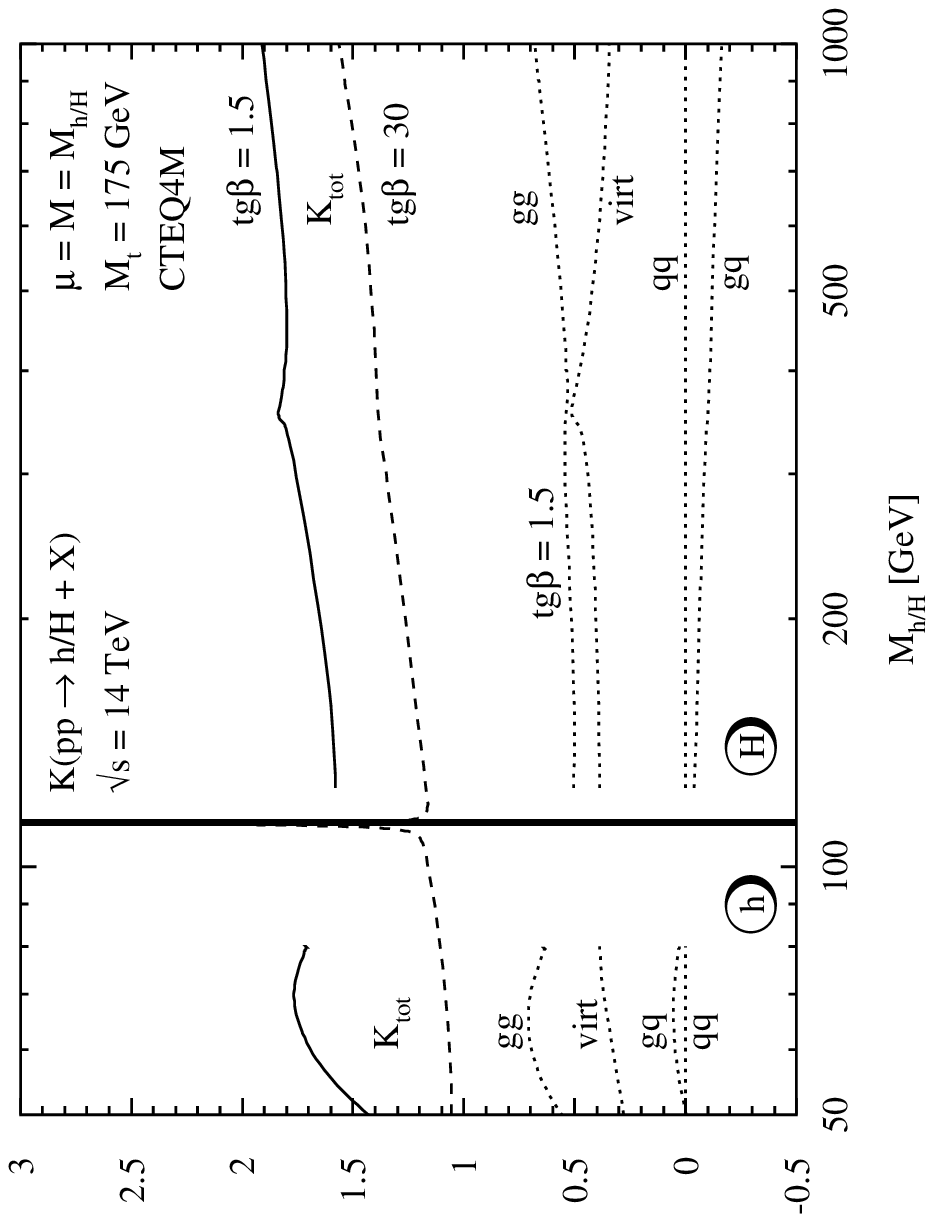}
\end{turn}
\vspace*{0.5cm}

\vspace*{-13.58cm}
\hspace*{2.5cm}
\begin{turn}{-90}%
\epsfxsize=9cm \epsfbox{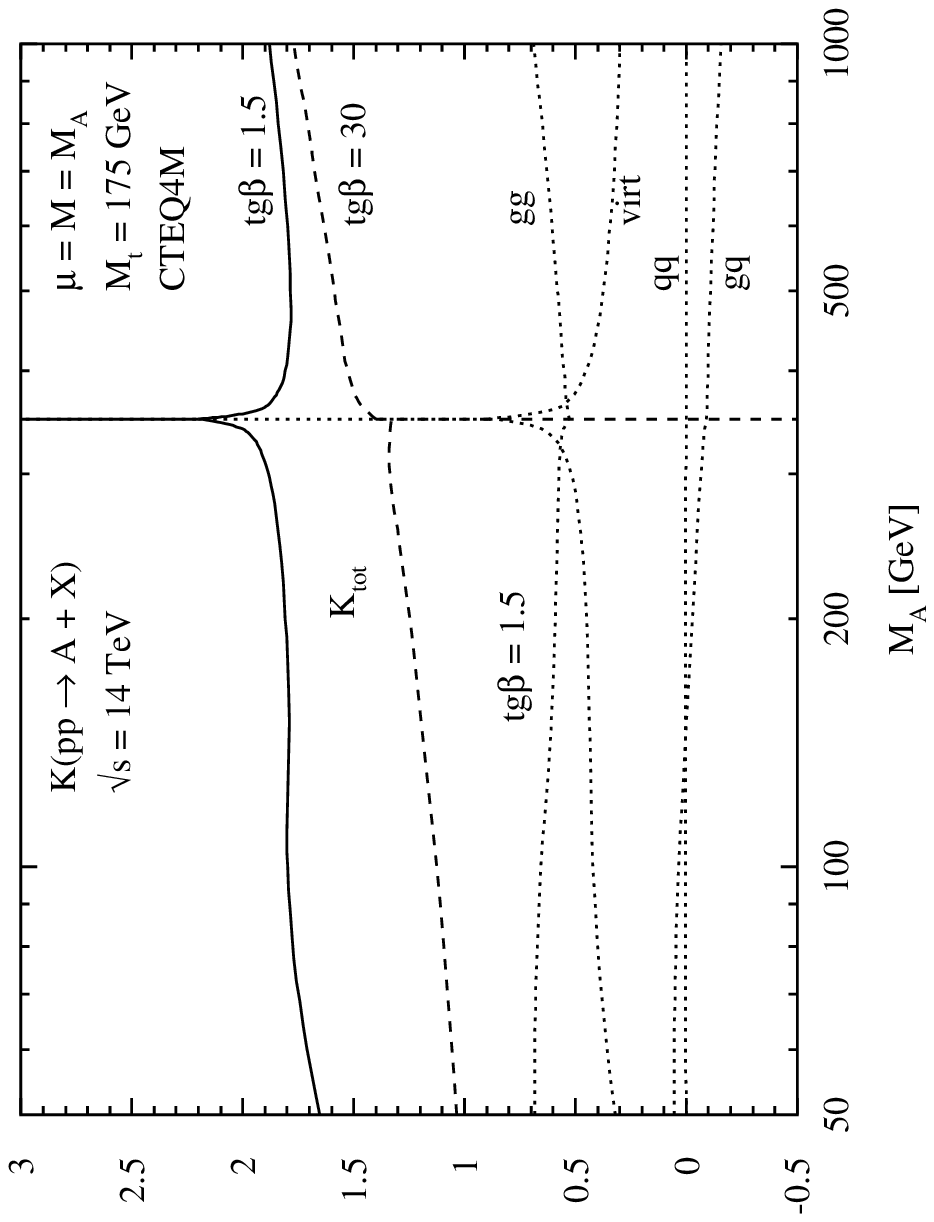}
\end{turn}
\vspace*{-0.8cm}

\caption[]{\label{fg:mssmgghk} \it K factors of the QCD-corrected gluon-fusion
cross section $\sigma(pp\to\phi+X)$ at the LHC with c.m.~energy $\sqrt{s}=14$
TeV. The dashed lines show the individual contributions of the four terms of
the QCD corrections given in eq.~(\ref{eq:signlo}). The renormalization and
factorization scales have been identified with the corresponding Higgs mass,
$\mu=M=M_\phi$, and the CTEQ4M parton densities have been adopted.}
\vspace*{-0.5cm}
\end{figure}
The NLO QCD corrections consist of two-loop virtual
corrections and one-loop real corrections from the processes
$gg\to Hg, gq\to Hq, q\bar q\to Hg$.
Typical diagrams are depicted in Fig.~\ref{fg:gghqcddia}. The evaluation of
the virtual and real corrections has been performed within dimensional
regularization. The five-dimensional Feynman
integrals of the two-loop diagrams have been reduced analytically to
one-dimensional ones, which have been integrated numerically.
The tensor reduction of the virtual three-point functions cannot
be performed completely down to scalar integrals, but one is left with the
calculation of irreducible tensor integrals. The heavy quark mass has
been renormalized on-shell and the strong coupling $\alpha_s$
in the \MS~scheme. The remaining collinear singularities of the real
corrections are absorbed in the NLO parton densities, defined
in the \MS~scheme. In order to calculate the QCD corrections to the
pseudoscalar Higgs boson production, a consistent scheme for the $\gamma_5$
coupling has to be used. We implemented the 't Hooft--Veltman scheme
\cite{thooft}, its modification by Larin \cite{larin} and the scheme
introduced by Kreimer \cite{kreimer}, all schemes giving identical results;
for a naive $\gamma_5$ the final result turned out to be ambiguous and
thus inconsistent, as a result of the ABJ anomaly \cite{abj}.
The cross sections can be cast into the
form \cite{gghqcd}
\bea
\sigma(pp \rightarrow \phi +X) & = & \sigma^\phi_{0} \left[ 1+ C^\phi
\frac{\alpha_{s}}{\pi} \right] \tau_\phi \frac{d{\cal L}^{gg}}{d\tau_\phi} +
\Delta\sigma^\phi_{gg} + \Delta\sigma^\phi_{gq} + \Delta\sigma^\phi_{q\bar{q}}
\label{eq:signlo} \\
C^\phi(\tau_Q) & = & \pi^{2}+ c^\phi(\tau_Q) +
\frac{33-2N_{F}}{6} \log \frac{\mu^{2}}{M_\phi^{2}} \non \\
\Delta \sigma^\phi_{gg} & = & \int_{\tau_\phi}^{1} d\tau \frac{d{\cal
L}^{gg}}{d\tau} \times \frac{\alpha_{s}}{\pi} \sigma^\phi_{0} \left\{ - z
P_{gg} (z) \log \frac{M^{2}}{\hat{s}} + d^\phi_{gg} (z,\tau_Q)
\right. \non \\
& & \left. \hspace{0.5cm} + 12 \left[ \left(\frac{\log
(1-z)}{1-z} \right)_+ - z[2-z(1-z)] \log (1-z) \right] \right\} \non \\ \non \\
\Delta \sigma^\phi_{gq} & = & \int_{\tau_\phi}^{1} d\tau \sum_{q,
\bar{q}} \frac{d{\cal L} ^{gq}}{d\tau} \times \frac{\alpha_{s}}{\pi}
\sigma^\phi_{0}\left\{ -\frac{z}{2} P_{gq}(z) \log\frac{M^{2}}{\hat{s}(1-z)^2}
\right. \non \\
& & \left. \hspace{2cm} + d^\phi_{gq} (z,\tau_Q) \right\}
\non \\ \non \\
\Delta \sigma^\phi_{q\bar{q}} & = & \int_{\tau_\phi}^{1} d\tau
\sum_{q} \frac{d{\cal L}^{q\bar{q}}}{d\tau} \times \frac{\alpha_{s}}{\pi}
\sigma^\phi_{0}~d^\phi_{q\bar q} (z,\tau_Q) \non
\eea
with $z = \tau_\phi / \tau = M_\phi^2/\hat s$. $P_{gg}$ and $P_{gq}$ are the
standard Altarelli--Parisi splitting functions \cite{AP}.
The
$K$ factors $K=\sigma_{NLO}/\sigma_{LO}$, where the NLO (LO) cross sections have
been evaluated with NLO (LO) $\alpha_s$ and parton densities, are presented
in Fig.~\ref{fg:mssmgghk} as functions of the corresponding Higgs masses.
The QCD corrections are large and positive, increasing the production cross
sections by 10--100\% \cite{gghqcd}, so that they can no
longer be neglected.
In spite of the large size of the QCD corrections the scale dependence is
reduced significantly, rendering these NLO results reliable within
$\sim 20\%$ \cite{gghqcd}, as can be seen in Fig.~\ref{fg:mssmgghscale}, which
shows the
LO and NLO cross sections as a function of the common renormalization and
factorization scales in units of the corresponding Higgs masses.
Improved measurements of the parton densities in the HERA experiments reduced
the uncertainties from the parton densities to a level
of $\sim 15\%$ \cite{gghqcd}. The size of the $K$ factors depends strongly on
$\tgb$ \cite{gghqcd}.

In the limit of heavy quark masses the mass-dependent terms of
Eq.~\ref{eq:signlo} reduce to very simple expressions \cite{gghqcd}:
\begin{eqnarray}
c^{h,H}(\tau_Q) & \displaystyle \to & \frac{11}{2}\hspace{1cm} c^A(\tau_Q)\to 6
\hspace{1cm} d_{gg}(z,\tau_Q) \to -\frac{11}{2} (1-z)^3
\non \\
d_{gq}(z,\tau_Q) \displaystyle & \to & \frac{2}{3}z^2 - (1 - z)^2 \hspace{2cm}
d_{q\bar q}(z,\tau_Q) \to \frac{32}{27} (1-z)^3 \label{eq:limit}
\end{eqnarray}
These limits can also be derived from low-energy theorems.

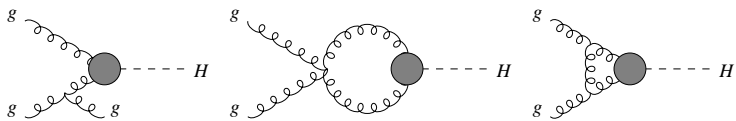
\begin{figure}[hbt]
\vspace*{-2cm}

\begin{center}
\SetScale{-0.6}
\SetPFont{Times-Italic}{12}
\begin{picture}(500,100)(-300,-50)

\DashLine(0,50)(50,50){5}
\Gluon(50,50)(75,65){3}{3}
\Gluon(75,65)(100,80){3}{3}
\Gluon(50,50)(75,35){-3}{3}
\Gluon(75,35)(100,20){-3}{3}
\Gluon(75,65)(75,35){3}{3}
\GCirc(50,50){10}{0.5}
\PText(-15,55)(-180)[lt]{H}
\PText(105,18)(-180)[lt]{g}
\PText(105,78)(-180)[lt]{g}

\DashLine(140,50)(190,50){5}
\GlueArc(215,50)(25,0,180){3}{8}
\GlueArc(215,50)(25,180,360){3}{8}
\Gluon(240,50)(290,80){3}{5}
\Gluon(240,50)(290,20){3}{5}
\GCirc(190,50){10}{0.5}
\PText(125,55)(-180)[lt]{H}
\PText(295,18)(-180)[lt]{g}
\PText(295,78)(-180)[lt]{g}

\DashLine(330,50)(380,50){5}
\Gluon(380,50)(405,65){3}{3}
\Gluon(405,65)(430,80){3}{2}
\Gluon(405,65)(380,80){-3}{2}
\Gluon(380,50)(430,20){-3}{5}
\GCirc(380,50){10}{0.5}
\PText(315,55)(-180)[lt]{H}
\PText(435,18)(-180)[lt]{g}
\PText(435,78)(-180)[lt]{g}
\PText(370,78)(-180)[lt]{g}

\end{picture}  \\
\setlength{\unitlength}{1pt}
\caption[]{\label{fg:effdia} \it Typical effective diagrams contributing to
the QCD corrections to $gg\to H$.}
\end{center}
\end{figure}

\paragraph{\it Low-energy theorems.}
For scalar Higgs bosons these rely on the fact that for
vanishing Higgs momentum the entire interaction of the Higgs particle with
fermions and gauge bosons can be generated by a shift of masses
by the Higgs field. Thus matrix elements with an external light Higgs
boson are related to the matrix element without the Higgs boson \cite{let}:
\begin{equation}
\lim_{p_H\to 0} {\cal M}_0(XH) = \frac{g^\phi}{v_0} m_0
\frac{\partial}{\partial m_0} {\cal M}_0(X)
\end{equation}
where $X$ denotes any particle configuration, $v=1/\sqrt{\sqrt{2}G_F}$ the
vacuum expectation value, and $g^\phi$ the corresponding SUSY coupling of
Table \ref{tb:susycoup}. In order to extend this relation to higher orders
in perturbation theory, it must be formulated in terms of bare quantities
\cite{gghqcd}.
For on-shell Higgs particles this mathematical limit coincides with the
massless Higgs limit. Thus we can derive the effective coupling of a light
Higgs boson to gluons from the gluon self-energy \cite{gghqcd},
\beq
{\cal L}_{eff} = \frac{\alpha_s}{12\pi}G^{a\mu\nu}G^a_{\mu\nu} \frac{H}{v}
\left\{1 + \frac{11}{4} \frac{\alpha_s}{\pi} + {\cal O}(\alpha_s^2)
\right\}
\label{eq:hggeff}
\eeq
This Lagrangian has to be interpreted as part of the basic Lagrangian
describing the effective theory, once the top quark is integrated out, and thus
provides the proper description of the heavy quark limit. The effective
coupling has to be inserted in the blobs of the effective diagrams shown in
Fig.~\ref{fg:effdia}. The final result coincides with
Eqs.~\ref{eq:signlo},\ref{eq:limit} of the scalar Higgs bosons.

The low-energy theorems for pseudoscalar Higgs particles are based on the ABJ
anomaly in the divergence of the axial-vector current \cite{abj},
\begin{equation}
\partial^\mu j_\mu^5 = 2m_Q \bar Q i\gamma_5 Q + \frac{\alpha_s}{2\pi}
G^{a\mu\nu} \widetilde{G}^a_{\mu\nu}
\label{eq:abj}
\end{equation}
with $\widetilde{G}^a_{\mu\nu} = \frac{1}{2} \epsilon_{\mu\nu\alpha\beta}
G^{a\alpha\beta}$
denoting the dual field strength tensor. The axial-vector current operator
fulfils the low-energy condition \cite{suthvel}
\beq
\langle gg| \partial_\mu j^\mu_5 A | A \rangle \to 0 \hspace{0.5cm} \mbox{for}
~p_A\to 0
\eeq
Using the basic interaction ${\cal L}_{int}=-g^A_Q m_Q \bar Q i\gamma_5 Q A/v$
the effective Lagrangian for the $Agg$ coupling can be derived
\cite{gghqcd}:
\begin{equation}
{\cal L}_{eff} = g_Q^A\frac{\alpha_s}{4\pi}G^{a\mu\nu}\widetilde{G}^a_{\mu\nu}
\frac{A}{v}
\end{equation}
Due to the Adler--Bardeen theorem of the non-renormalization of the ABJ anomaly
\cite{adbar}, this effective Lagrangian is valid up to all orders of
perturbation theory.
Its insertion into the effective diagrams analogous to Fig.~\ref{fg:effdia}
leads to the results of Eqs.~\ref{eq:signlo},\ref{eq:limit} of the pseudoscalar
Higgs boson.

\begin{figure}[hbt]

\vspace*{-0.9cm}
\hspace*{3.0cm}
\epsfxsize=5cm \epsfbox{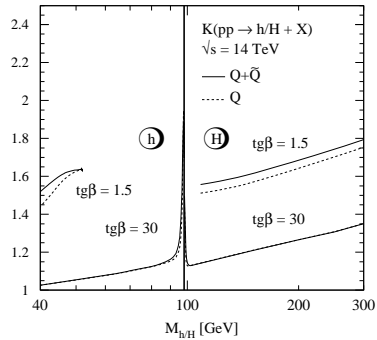}
\vspace*{-1.7cm}

\caption[]{\label{fg:kqksq} \it K factors of the cross sections $\sigma(pp\to
h/H + X)$ with [solid lines] and without [dashed lines] squark loops as a
function of the corresponding scalar Higgs mass for two values of $\tgb=1.5,30$.
The common squark mass has been chosen as $M_{\widetilde{Q}}=200$ GeV.}
\vspace*{-0.5cm}
\end{figure}

\paragraph{\it Squark loops.}
If the full massive LO cross section is multiplied by the $K$ factor obtained
in the heavy quark limit, a reliable approximation is obtained, within 10\%
for the top quark contribution to the production cross sections \cite{habil}.
Thus it will be sufficient to derive the QCD corrections to squark
loops in the heavy squark limit. This can be done by extending the scalar
low-energy theorems to squarks. We performed this calculation in the
approximation of degenerate squark flavors and very heavy gluinos so that the
latter decouple. The effective Lagrangian mediated by a squark loop reads
\cite{sqcd}
\beq
{\cal L}_{eff} = \frac{\alpha_s}{48\pi}G^{a\mu\nu}G^a_{\mu\nu} \frac{H}{v}
\left\{1 + \frac{25}{6} \frac{\alpha_s}{\pi} + {\cal O}(\alpha_s^2)
\right\}
\label{eq:aggeff}
\eeq
The amplitudes can be obtained from effective diagrams in analogy
to Fig.~\ref{fg:effdia}. They were added
to the massive quark amplitudes and contracted with the massive LO form factors
in the virtual corrections. In this way we derived the most reliable
approximation. The $K$ factors differ only by less than
$\sim 10\%$ from the $K$ factors for the quark loops alone \cite{sqcd}. This is
shown in
Fig.~\ref{fg:kqksq}, which presents the $K$ factors with and without squark
loops for a common squark mass of 200 GeV, for which the LO cross sections are
significantly enhanced. Thus the full $K$ factors can simply be approximated
by the massive $K$ factors to the quark loops alone, while the LO cross sections
should include the massive squark contributions.
This approximation is valid, because the QCD corrections for heavy particle
loops are dominated by soft and collinear gluon effects.

\subsection{Soft gluon resummation}
\begin{figure}[hbt]

\vspace*{0.0cm}
\hspace*{-0.2cm}
\begin{turn}{-90}%
\epsfxsize=4cm \epsfbox{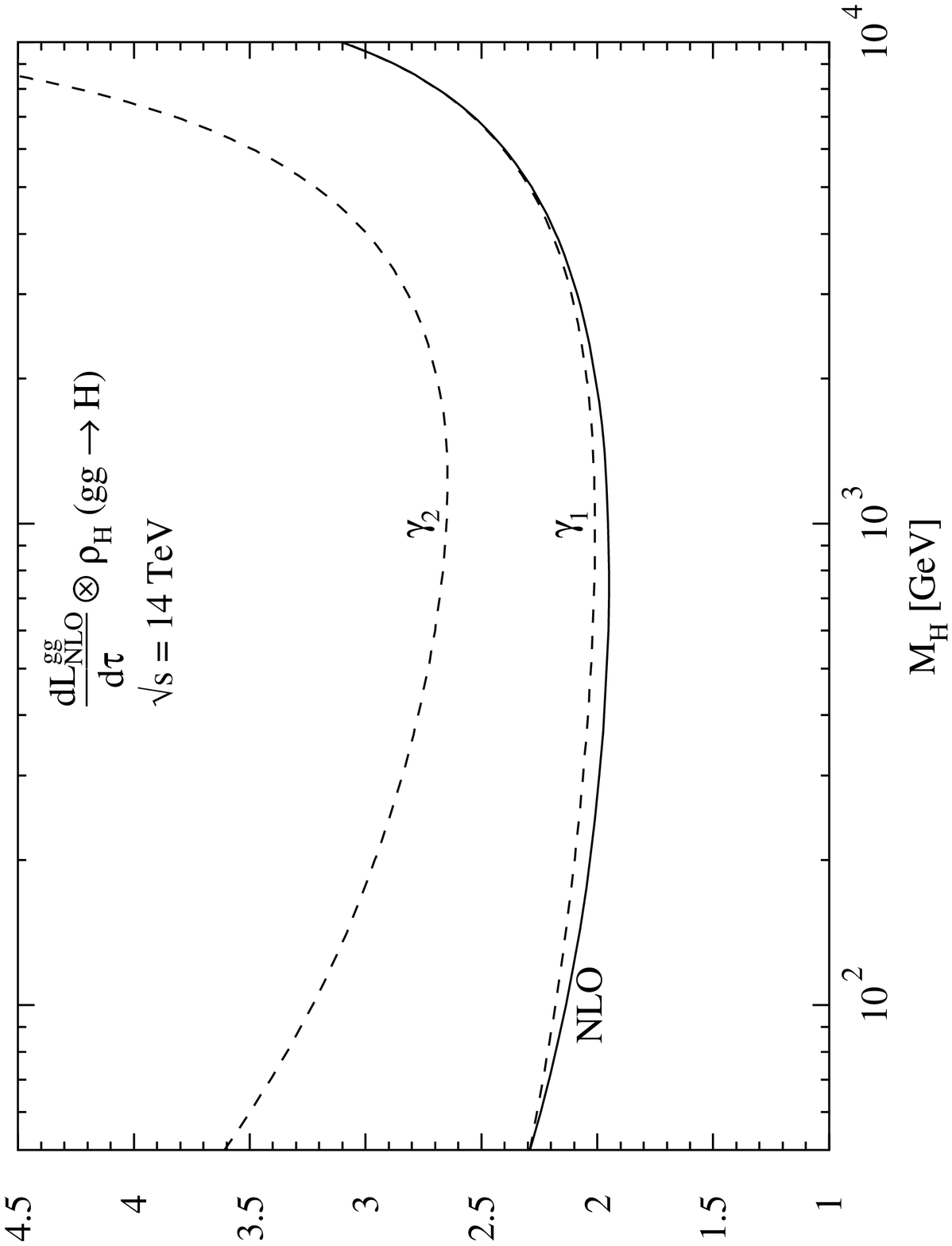}
\end{turn}
\vspace*{-0.0cm}

\vspace*{-4.03cm}
\hspace*{5.8cm}
\begin{turn}{-90}%
\epsfxsize=4cm \epsfbox{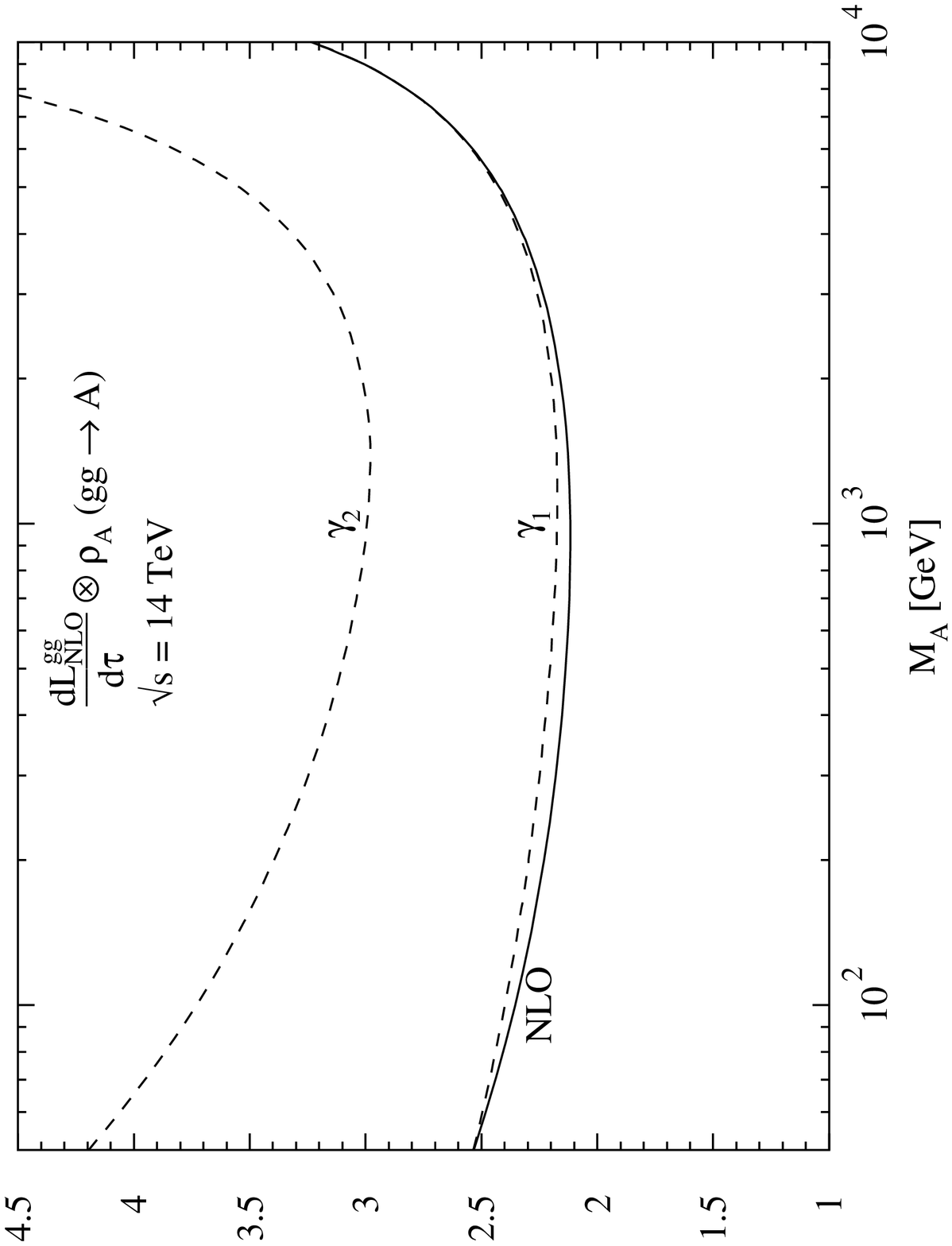}
\end{turn}
\vspace*{-0.1cm}

\caption[]{\label{fg:resum} \it Exact and approximate two- and three-loop
correction factor convoluted with NLO gluon densities in the heavy top quark
limit for the MSSM Higgs bosons. The CTEQ4M parton densities have been adopted
with $\alpha_s(M_Z)=0.116$ at NLO.}
\vspace*{-1.0cm}
\end{figure}

The results of the last paragraph provide a strong motivation for the
resummation of soft gluon effects in the Higgs boson production cross
sections. This step has been carried out for a common squark mass
$m_{\squ} = 1$ TeV, so that squark loops
can be neglected \cite{sqcd}. Moreover, we worked in the heavy quark limit,
which provides
an approximation to the cross sections within $\sim 25\%$ for $\tgb \lsim 5$.
For larger values of $\tgb$ the bottom contribution becomes significant. Finally
the $gq$ and $q\bar q$ initial states will be neglected, since their
contribution only amounts to $\sim 10\%$ \cite{gghqcd}. Within this
approximation the
partonic cross sections factorize into three pieces \cite{resum},
\beq
\hat\sigma_{gg} = \sigma_0^\phi \kappa_\phi \rho_\phi
\eeq
The factors $\kappa_\phi$ are fixed by the effective Lagrangian at NNLO
\cite{resum,CKS}:
\bea
\kappa_A & = & 1 \\
\kappa_{h,H} & = & 1 + \frac{11}{2} \frac{\alpha_s(M_t)}{\pi}
+ \frac{3866 - 201\, N_F}{144} \left(\frac{\alpha_s(M_t)}{\pi} \right)^2
\nonumber \\
& & \hspace*{1cm} + \frac{153-19 N_F}{33-2 N_F}
~\frac{\alpha_s(M_{H}) - \alpha_s(M_t)}{\pi}
+ {\cal O}(\alpha_s^3) \non
\eea
while the correction factors $\rho_\phi$ originate from effective diagrams.
At LO they are normalized as $\rho_\phi=\delta(1-z)$ with $z=M_\phi^2/\hat s$.
It has been shown for the Drell--Yan process that
for $z\lsim 1$ the cross section factorizes into soft gluon contributions,
jet functions containing collinear gluon effects and a hard matrix element
\cite{factheo}.
From this factorization the Sudakov evolution equation can be derived
\cite{laenen}:
\beq
M_\phi^2 \frac{\partial \rho_\phi}{\partial M_\phi^2} = W_\phi \otimes\rho_\phi
\label{eq:sud}
\eeq
where the convolution is defined as $f\otimes g = \int_z^1 dz' f(z') g(z/z')$.
The evolution kernel $W_\phi$ can be determined from the perturbative result
by matching the leading terms with the perturbative expansion of the solution to
Eq.~\ref{eq:sud}. In this way we have determined the first term in the
expansion of $W_\phi$ and thus resummed the soft gluon effects with NLO
accuracy. The renormalization and mass factorization defined the strong
coupling $\alpha_s$ and the parton densities in the \MS~scheme. The resummed
result has then been used as a generating functional for the perturbative
expansion up to NNLO. The NLO terms of $\rho_\phi$ read as \cite{resum}
\bea
\rho^{(1)}_{h,H} & = & 12{\cal D}_1(z) - 24{\cal E}_1(z) - 6 {\cal D}_0(z) L_\mu
+ \pi^2 \delta(1-z) \\
\rho_A^{(1)} & = & \rho_{h/H}^{(1)} + 6 \delta(1-z)
\eea
where
\beq
{\cal D}_i(z) = \left[\frac{\log^i(1-z)}{1-z}\right]_+ \, ,
\quad \quad
{\cal E}_i(z) = \log^i(1-z) \, ,
\quad \quad
 L_\mu = \log \left( \frac{\mu^2}{M_\phi^2} \right)
\eeq
The term ${\cal E}_1(z)$ extends the conventional resummation techniques,
which only
resum the soft gluon logarithms ${\cal D}_i(z)$. The leading part of the
${\cal E}_i(z)$ terms
is of a pure collinear nature and thus universal as well. The
consistency of this extension has not been proved so far. However, it is
important to note that these terms are large for LHC processes, and thus
relevant.
Moreover, the analogous analysis for the Drell--Yan process and deep inelastic
scattering reproduced the leading terms in ${\cal E}_i(z)$ at the NNLO level
so that
we got confidence that they are universally factorizing in the collinear limit.
The NNLO expansion of $\rho_\phi$ can be cast into the form
\cite{resum}
\bea
\rho^{(2)} & = &
3\left\{
      24 {\cal D}_3(z) +
      (-2 \beta_0 - 36 L_\mu ) {\cal D}_2(z)
    + ( - 24 \zeta_2  +2 \beta_0 L_\mu  +
            12 L_\mu^2) {\cal D}_1(z) \right. \nonumber \\
&+&
      (48 \zeta_3 + 12 \zeta_2 L_\mu -
            \frac{1}{2} \beta_0 L_\mu^2){\cal D}_0(z)
       - 48 {\cal E}_3(z)
\nonumber \\
&+&
       (4\beta_0 + 24 + 72 L_\mu ){\cal E}_2(z)
  + (48 \zeta_2-4\beta_0 L_\mu - 24 L_\mu - 24 L_\mu^2){\cal E}_1(z)
\nonumber \\
&+& \left.
    (18 \zeta_2^2 - 36 \zeta_4 - \frac{2909}{432}\beta_0
               + \zeta_2 \beta_0 L_\mu
          - 24 \zeta_3L_\mu -
              6 \zeta_2 L_\mu^2 ) \delta(1-z)  \right\} \non \\
\rho_A^{(2)} & = & \rho_{h/H}^{(2)}
+ 3\left\{ 24 {\cal D}_1(z) - 12 L_\mu {\cal D}_0(z) - 48 {\cal E}_1(z)
\right. \nonumber \\
& & \left. + ( 12 \zeta_2 + 6 + \beta_0 L_\mu) \delta(1-z) \right\}
\label{eq:resum}
\eea
with $\beta_0 = (33-2N_F)/6$.
The analogous analysis of the Drell--Yan
process results in a reliable approximation up to NNLO \cite{resum} so that
the NNLO
results of Eq.~\ref{eq:resum} are expected to be a reliable approximation of
the NNLO corrections to the Higgs production cross sections. The correction
factors convoluted with NLO parton densities are presented in
Fig.~\ref{fg:resum} as a function of the corresponding Higgs mass. The NLO
expansion of $\rho_\phi$ is denoted by $\gamma_1$ and the NNLO expansion by
$\gamma_2$.
We observe a good agreement of $\gamma_1$ with the exact NLO result, within
$\sim 5\%$, while
the correction factor $\gamma_2$ turns out to be large \cite{resum}. This,
however, is
caused by the use of NLO quantities in all perturbative orders of the
correction factor. A consistent analysis requires LO $\alpha_s$ and parton
densities for the LO correction factor and NNLO quantities for $\gamma_2$.
This reduces the NLO correction factor to a level of $\sim 1.5$, which is
significantly smaller than the NLO and $\gamma_1$ curves in Fig.~\ref{fg:resum}.
Thus in order to obtain a NNLO prediction NNLO parton densities are needed,
which, however,
are not available so far. Therefore a consistent theoretical prediction of the
NNLO Higgs production cross sections is not yet possible.

\section{Conclusions}
We have reviewed the theoretical status of the production of neutral MSSM
Higgs bosons at the LHC. The NLO QCD corrections are large and
positive, while the theoretical uncertainty estimated from the residual scale
dependence is reduced significantly. The squark loop contributions become
sizeable for squark masses below $\sim 400$ GeV. The $K$ factors, however,
deviate by less than $\sim 10\%$ if
squark loops are included. The $K$ factors depend strongly on the
MSSM parameter $\tgb$. Finally we have performed the soft gluon resummation for
the Higgs production cross sections, which we expanded up to NNLO in order to
get a reliable estimate of the NNLO corrections. These turn out to be
potentially large. However, a consistent phenomenological analysis at NNLO
requires NNLO parton densities, which are not yet available.

\section*{Acknowledgements}
I would like to thank S.\ Dawson, A.\ Djouadi, D.\ Graudenz, M.\ Kr\"amer,
E.\ Laenen and P.\ Zerwas for the fruitful collaboration in the presented work.

\end{document}